# Visualization of Wearable Data and Biometrics for Analysis and Recommendations in Childhood Obesity

Michael Aupetit, Luis Fernandez-Luque, Meghna Singh and Jaideep Srivastava
*Qatar Computing Research Institute (QCRI)*
*Hamad Bin Khalifa University, Doha, Qatar*
Email: {maupetit, lluque, mesingh, jsrivastava}@hbku.edu.qa

*Abstract*—Obesity is one of the major health risk factors behind the rise of non-communicable conditions. Understanding the factors influencing obesity is very complex since there are many variables that can affect the health behaviors leading to it. Nowadays, multiple data sources can be used to study health behaviors, such as wearable sensors for physical activity and sleep, social media, mobile and health data. In this paper we describe the design of a dashboard for the visualization of actigraphy and biometric data from a childhood obesity camp in Qatar. This dashboard allows quantitative discoveries that can be used to guide patient behavior and orient qualitative research.

## I. INTRODUCTION

Childhood obesity is a growing epidemic, and with technological advancements, new tools can be used to monitor and analyze lifestyle factors leading to obesity, which in turn can help in timely health behavior modifications. In this paper we present a tool for visualization of personal health data, which can assist healthcare professionals in designing personalized interventions for improving health. The data used for the tool was collected as part of a research project called "Adaptive Cognitive Behavioral Approach to Addressing Overweight and Obesity among Qatari Youth" (ICAN Study)[1]. The participants in the study were involved in activities aimed at improving their health behavior and losing weight. All participants and their parents/guardians provided informed consent prior to participation. Data from various sources (social media, mobile, wearables and health records) were collected from subjects and linked using a unique subject identifier. These datasets provided what we have defined as a 360-degree Quantified Self (360QS) view of individuals [2], [3].

We have focused on the visualization of the biometrics and physical activity data. We discuss different visualization techniques used to analyze the activity patterns of participants in the obesity trial. Our dashboard is designed to compare data across time, and among reference individuals.

## II. IMPLEMENTATION OF THE DASHBOARD

### A. Data Handling

Biometric data were measured periodically and included height, weight and the derived body-mass index (BMI), body fat percentage, waist circumference and blood pressure for each individual. Physical activity data was collected via accelerometers. The raw signals have been quantized into four activity levels: sedentary, light, moderate and vigorous, using a human activity recognition algorithm [4].

### B. Visualization

The objective of the dashboard is to provide an overview of the actigraphy data and enable primary data exploration by an expert user. Referring to figure 1, the interface is made up of a *Control Panel* (left side) and a *Visualization Panel* (right side).

In the *Control Panel*, drop down menus enable selecting two subjects to be compared based on identifiers and gender; e.g., boys number 84 (top) and 82 (bottom) are displayed. During data collection, some devices were not worn at all times; hence they recorded long periods of "sedentary" activity. The user can use a slider to filter out days with more than the selected number of sedentary hours. Finally, the user can use check-boxes to select the biometrics she wants to compare (e.g., BMI and Body Fat Percentage).

The *Visualization Panel* shows both subjects, one at the top in a box with orange frame and the other at the bottom in a box with violet frame. The top and bottom boxes show a bar chart indicating the hours of activity (y-axis) with activity level breakdown per day through time (x-axis). The color legend of the bars is shown in the control panel: reddish colors for moderate (light red) and vigorous (dark red) activity levels, and bluish colors for light (light blue) and sedentary (dark blue) activity levels. The user can select a time window by brushing horizontally on the time range to zoom in, and double clicking the background to zoom out and see the entire time period. The line charts at the center show the biometrics selected in the control panel, with line colors corresponding to the colored frame (orange line for subject displayed at the top, violet for the one at the bottom). The biometrics have not been measured on a daily basis, which explains the scarce number of measurements linearly interpolated here. On the right side, the average activity breakdown by activity levels is displayed for weekdays (top and bottom) and for weekend days (middle).

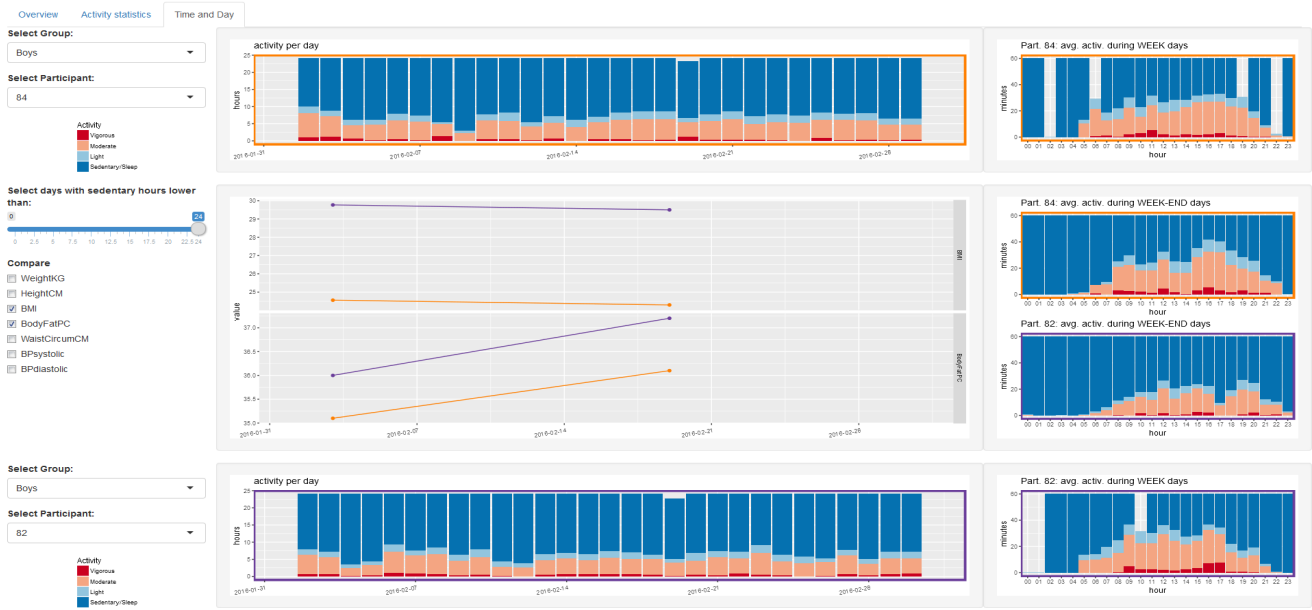

Figure 1. The visualization dashboard for actigraphy data and biometrics analysis.

*C. Quantitative Analysis based on Visualization*

Referring to figure 1, subjects 84 (top) and 82 (bottom) have similar activity level patterns through the time period selected. However subject 82 has a higher BMI (violet line) than subject 84. Both subjects had a body fat increase over time. The difference between their weights is correlated to their week and weekend activity levels: subject 84 is more active during the weekend than subject 82.

## III. CONCLUSIONS

This interface must be seen as a tool to give primary overview of the data likely to orient more detailed analysis. For instance, a more in-depth study of the relation between sleep duration and BMI could be conducted. Another outcome related to the experimental setup would consist of recommending biometrics to be measured more often, or to find incentives for subjects to wear the devices more consistently.

This dashboard could be used to assess a subject's health status against the group's instead of that of another subject. A health expert could also provide the subject with a target status (e.g. weight) to compare and converge to, along with recommendations about the activities he/she should improve: e.g., go to bed earlier, wake up earlier during weekend, have more vigorous activities during afternoon, etc.

Other available tools, such as the Fitbit dashboard (Fitbit Inc, USA), do not give detailed activity levels across time nor comparison with reference individual. Other ones [5] focus on the representation of the data rather than its exploration and companionship across the groups. Our next steps include performing qualitative evaluation of our dashboard and improvements based on the end users' feedback.

ACKNOWLEDGMENT

We would like to acknowledge the contributions of Dr. Mohamed Ahmedna from Qatar University, and all the members of the ICAN project, along with our colleagues at QCRI. The ICAN project was co-funded by the Qatar National Research Fund (a member of Qatar Foundation) under project number NPRP X- 036- 3- 013.